\documentstyle[aps,prl,epsf,graphicx,multicol]{revtex}

\begin{document}
\renewcommand{\narrowtext}{\begin{multicols}{2} \global\columnwidth20.5pc}
\renewcommand{\widetext}{\end{multicols} \global\columnwidth42.5pc}

\def\tende#1{\,\vtop{\ialign{##\crcr\rightarrowfill\crcr
\noalign{\kern-1pt\nointerlineskip}
\hskip3.pt${\scriptstyle #1}$\hskip3.pt\crcr}}\,}

\title{Coherence length in superconductors from weak to strong coupling}
\author{ L. Benfatto, A. Toschi, S. Caprara,  and  C. Castellani}

\address{Dipartimento di Fisica, Universit\`a di Roma ``La Sapienza'',\\  
         and Istituto Nazionale per la Fisica della Materia (INFM), 
         SMC and Unit\`a di Roma 1, \\
         Piazzale Aldo Moro, 2 - 00185 Roma, Italy}
\maketitle
\begin{abstract}
We study the evolution of the superconducting coherence length 
$\xi_0$ from weak to strong coupling, both within a $s$-wave and a $d$-wave 
lattice model. We show that the identification of $\xi_0$ with the 
Cooper-pair size $\xi_{pair}$ in the weak-coupling regime is meaningful only 
for a fully-gapped (e.g., $s$-wave) superconductor. Instead in the $d$-wave 
superconductor, where $\xi_{pair}$ diverges, we show that $\xi_0$ is properly 
defined as the characteristic length scale for the correlation function of the 
modulus of the superconducting order parameter. The strong-coupling regime is 
quite intriguing, since the interplay between particle-particle and 
particle-hole channel is no more negligible. In the case of $s$-wave pairing, 
which allows for an analytical treatment, we show that $\xi_0$ is of order 
of the lattice spacing at finite densities. In the diluted regime $\xi_0$ 
diverges, recovering the behavior of the coherence length of a weakly 
interacting effective bosonic system. Similar results are expected to hold 
for $d$-wave superconductors.  

\end{abstract}
~\par\noindent
PACS numbers: 74.20.-z,74.20.Fg,71.10.Fd
\begin{multicols}{2} \global\columnwidth20.5pc

\section{Introduction}
In the last years the analysis of the phase-fluctuation contribution to the 
low-temperature properties of high-$T_c$ superconducting cuprates stimulated 
the interest in the microscopic derivation of the phase-only action for 
$d$-wave superconductors \cite{randeria,noi,sharapov,noi2}. The symmetry of 
the order parameter has been shown to play a crucial role in determining  the 
quantum-to-classical crossover for phase fluctuations via enhanced 
dissipative effects \cite{randeria,noi}. The effective hydrodynamic 
phase-only action is  usually cut off at large momenta. A reasonable spatial 
bound for the phase-only action is the coherence length $\xi_0$, which sets 
the distance above which the fluctuations of the modulus of the order 
parameter $|\Delta|$ become uncorrelated. Indeed, at distances greater than 
$\xi_0$ the relevant degrees of freedom are the phase fluctuations only. 
Within this context, a proper definition of $\xi_0$ must be related to the 
characteristic length scale of the spatial decay of the correlation function 
for $|\Delta|$, whatever is the symmetry of the order parameter.  

Another relevant length scale for superconductors is the characteristic size 
of the Cooper pair
\begin{equation}
\xi_{pair}=\sqrt{\frac{\int d {\bf r}\, |\psi({\bf r})|^2 {\bf r}^2}
{\int d {\bf r}\, |\psi({\bf r})|^2}}=\sqrt{
\frac{\int d {\bf k} \,[{\partial_{\bf k}} \phi({\bf k})]^2}{\int d {\bf k} 
\,\phi^2({\bf k})}},
\label{csipair}
\end{equation}
where $\psi({\bf r})$ is the Cooper-pair wave function, ${\partial_{\bf k}}$
is the gradient operator in ${\bf k}$-space, and 
$\phi({\bf k})=\Delta_0\gamma_{\bf k}/E_{\bf k}$ is the Fourier transform of 
$\psi({\bf r})$ \cite{libro}. Here $\Delta_0$ is the amplitude of the 
superconducting gap, $\gamma_{\bf k}$ is the factor which controls the 
symmetry of the BCS wave function, 
$E_{\bf k}\equiv\sqrt{\xi_{\bf k}^2+\Delta_0^2\gamma_{\bf k}^2}$, and 
$\xi_{\bf k}$ is the normal-state dispersion. Throughout this paper we assume 
the lattice spacing $a$ as the unit length. 

It is known that at weak coupling for a $s$-wave superconductor (i.e., 
$\gamma_{\bf k}=1$) both $\xi_0$ and $\xi_{pair}$ are, within a numerical 
factor, of the same order $v_F/\Delta_0$ \cite{thouless}. Here $v_F$ is a 
properly defined average value of the quasiparticle velocity at the Fermi 
surface (see Sec. II), which reduces to $k_F/m$ in a continuum model. By 
analogy, one would expect the same result to hold for $d$-wave 
superconductors ($\gamma_{\bf k}=\gamma_{\bf k}^d\equiv\cos k_x-\cos k_y$). 
However, when evaluating Eq. (\ref{csipair}), one finds that the mean size of 
the Cooper pair is {\em infinite}. This fact, which has been often overlooked 
(see, however, Ref. \cite{salkola}), can be easily seen by evaluating the 
dominant contribution to Eq. (\ref{csipair}), which comes from the nodal 
quasiparticles. As we show in Appendix A, the nodal quasiparticles give a 
logarithmically divergent contribution to Eq. (\ref{csipair}). The question 
then arises whether $\xi_0$ is well defined in the case of $d$-wave pairing, 
and of the expected order of magnitude in the weak-coupling limit. 

The strong-coupling limit is even more intriguing, as one would expect that 
the magnitudes of the two lengths are widely separated. For instance, in the 
case of $s$-wave pairing, while $\xi_{pair}$ should decrease as the pairing 
strength increases, $\xi_0$ should resemble the behavior of the natural 
length scale for the spatial variation of the condensate wave function of an 
effective ``bosonic'' system \cite{fetter,nozieres}. Thus $\xi_0$ should be 
controlled by the inverse of the residual interaction between the bosonic 
Cooper pairs, and could eventually increase as the residual boson-boson 
interaction decreases and by approaching the diluted limit. Such a separation 
between the two length scales has indeed been found in Ref. \cite{pistolesi} 
within the continuum $s$-wave model {\em \`a la} Nozi\`eres and Schmitt-Rink 
\cite{nsr}. 
 
In this paper we address the issue of the systematic analysis of the behavior 
of the coherence length $\xi_0$ defined through the spatial decay of the 
static correlation function for $|\Delta|$, $X_\Delta({\bf r})$, which is
obtained by Fourier-transforming the correlation function in momentum space
$X_\Delta({\bf q})$. We specifically consider the two-dimensional 
negative-$U$ Hubbard model for $s$-wave pairing and its simplest 
generalization for $d$-wave pairing. 

We show that, whatever is the symmetry of the order parameter, 
in the weak-coupling limit $\xi_0$ is finite and of the expected order 
$v_F/\Delta_0$. 

In the strong-coupling limit the modulus of the order parameter $|\Delta|$ 
and the density of particles $\rho$ are coupled. Specifically, in the case of
$s$-wave pairing, they experience the same fluctuations: in particular at low 
density the two fields become proportional \cite{micnas,depalo}. As a 
consequence, the density fluctuations contribute to $X_\Delta({\bf q})$ via 
the gap-density coupling. We derive $X_\Delta({\bf q})$ in $s$-wave 
superconductors by including density fluctuations (whose contribution to 
$\xi_0$ is negligible at weak coupling) and investigate the length scale of 
the spatial decay of its Fourier transform $X_\Delta({\bf r})$ by varying the 
density $\rho$ between 0 and 1 (the range $1\le \rho \le 2$ being recovered 
by particle-hole symmetry). In the diluted limit ($\rho\approx 0$) 
$X_\Delta({\bf r})$ decays exponentially with a length scale $\xi_0$  
diverging as $1/\sqrt \rho$, as it is expected for a weakly-coupled diluted 
Bose liquid \cite{micnas,fetter,nozieres}. At higher densities, and for strong
coupling, $X_\Delta({\bf r})$ oscillates with the periodicity $a$ of the 
lattice. This is due to the fact that $X_\Delta({\bf q})$ is dominated by 
momenta ${\bf q}\approx{\bf Q}\equiv(\pi,\pi)$, in analogy with the density 
mode, which is massless for ${\bf q}={\bf Q}$ at half filling \cite{cdw}. By
means of numerical calculations we show that the length above which 
$X_\Delta({\bf r})$ is strongly suppressed with respect to its value at 
${\bf r}=0$ is of the order of the lattice constant for all densities away 
from $\rho\approx 0$, even though it exhibits a long-living exponential tail 
governed by an increasing characteristic length scale approaching half filling.

The case of $d$-wave pairing in the strong-coupling limit would be much more 
difficult to address, as the coupling of the gap fluctuations to the 
particle-hole channel reflects in a Hartree-Fock-like correction to the bare 
band dispersion, making the analytical treatment not viable. Despite these 
complications, however, the physics is not expected to be different from the 
$s$-wave case, which allows for a much simpler and transparent analytical 
treatment. For this reason, we devote most of our strong-coupling analysis to 
the $s$-wave case.

Finally, we comment on the consequences of such an analysis on the problem
of a proper definition of the momentum cutoff for phase fluctuations in the 
effective phase-only model.

The plan of the paper is the following. In Sec. II we discuss the 
weak-coupling regime for both $s$- and $d$-wave superconductors. In Sec. III
we devote our analysis to the strong-coupling regime for $s$-wave 
superconductors. In particular, Sec. III A addresses the low-density regime,
Sec. III B deals with the intermediate- and high-density regime, while in
Sec. III C we discuss the properties of the density-density correlation
function. Concluding remarks are found in Sec. IV. Detailed calculations of 
the coherence length and of the Cooper-pair size for both $s$- and $d$-wave 
superconductors at weak coupling are reported in Appendix A. In Appendix B
we discuss how the strong-coupling results discussed in Sec. III are affected
by the inclusion of a next-to-nearest-neighbor hopping term.
 
\section{The weak-coupling regime}

We start with the BCS action on a two-dimensional lattice, at a temperature 
$T=\beta^{-1}$: 
\begin{eqnarray}
{\cal S}&=&\int_0^\beta d\tau \left\{\sum_{{\bf k},\sigma} 
c^+_{{\bf k} \sigma}(\tau) (\partial_\tau+\xi_{\bf k})
c_{{\bf k}\sigma}(\tau)+H_I(\tau)\right\},\nonumber\\
H_I&=&-\frac{U}{N_s}\sum_{{\bf k},{\bf k}',{\bf q}}
\gamma_{\bf k}\gamma_{{\bf k}'}c^+_{{\bf k}+\frac{{\bf q}}{2} \uparrow}
c^+_{-{\bf k}+\frac{{\bf q}}{2}  \downarrow}
c_{-{\bf k}'+\frac{{\bf q}}{2} \downarrow} 
c_{{\bf k}'+\frac{{\bf q}}{2}  \uparrow}.
\label{mod}
\end{eqnarray}
Here $N_s$ is the number of lattice sites, $U>0$ is the pairing interaction 
strength, $\xi_{\bf k}=-2t(\cos k_x+\cos k_y)-\mu$ is the band dispersion 
associated with a nearest-neighbor hopping $t$, $\mu$ is the chemical 
potential, and the factor $\gamma_{\bf k}$ controls the symmetry of the gap. 
To derive the correlation function for $|\Delta|$, we first perform the 
standard Hubbard-Stratonovich decoupling of $H_I$ and then make explicit the 
dependence on the phase $\theta$ and on the modulus $|\Delta |$ of the 
complex order parameter $\Delta=|\Delta|{\rm e}^{i\theta}$ by means of the 
gauge transformation $c_\sigma \rightarrow c_\sigma {\rm e}^{i\theta/2}$ 
\cite{randeria,sharapov,depalo}. As a consequence, the action depends on 
$|\Delta|$ through the Hubbard-Stratonovich Gaussian term and the interaction 
of the field with the fermions, which arises from $H_I$, i.e.
\begin{eqnarray*}
&~&{\cal S}(|\Delta|)=\sum_{q}\frac{1}{U}|\Delta|_q|\Delta|_{-q}\nonumber\\
&~&-\frac{T}{N_s}\sum_{k,k'}
\gamma_{\frac{{\bf k}+{\bf k}'}{2}}\Psi^+_{k'}\left(\begin{array}{cc}
0 & |\Delta|_{k'-k} \\|\Delta|_{k'-k} & 0 \end{array} \right)\Psi_k,
\nonumber
\end{eqnarray*}
where $q=({\bf q},\Omega_m)$, $k=({\bf k},\omega_n)$, $\Omega_m$ and
$\omega_n$ are the bosonic and fermionic Matsubara frequencies respectively, 
and we introduced the Nambu spinor 
$\Psi^+_k=(c_{k,\uparrow}^+,\,c_{-k, \downarrow})$. We assume, as usual, 
small fluctuations of $|\Delta|_q$ around its saddle-point value $\Delta_0$, 
i.e., $|\Delta|_q=\Delta_0+\delta|\Delta|_q$. After integrating out the 
fermionic degrees of freedom around the BCS superconducting saddle-point 
solution, we expand the resulting effective action ${\cal S}_{eff}$ for 
$\delta|\Delta|_q$ up to the second order, obtaining 
\begin{eqnarray} 
\label{effact}
{\cal S}_{eff}(\delta|\Delta|) & = & 
\sum_{q}\left[\frac{1}{U} -\frac{1}{2} D(q)\right]
\delta|\Delta|_q\delta|\Delta|_{-q}, \\ 
D({\bf q},\Omega_m) & = &  \frac{T}{N_s}\sum_{{\bf k}, \omega_n} 
\gamma_{\bf k}^2 \mbox{Tr} 
\left[{\cal G}_0\left({\bf k}+\frac{\bf q}{2},\omega_n+\Omega_m\right) 
\right.\nonumber\\
& \times&\left.
\tau_1{\cal G}_0\left({\bf k}-\frac{\bf q}{2},\omega_n\right)\tau_1
\right],\nonumber
\end{eqnarray}
where ${\cal G}_0$ is the Nambu Green function evaluated at the BCS level,
$\tau_1$ is the Pauli matrix and the trace is taken in the Nambu space.

We define
\begin{equation}
X_{\Delta}({\bf q},\Omega_m)=2<\delta|\Delta|_q \delta|\Delta|_{-q}>,
\label{defxd}
\end{equation}
i.e., the inverse of the coefficient of the Gaussian term in Eq. 
(\ref{effact}). Since we are interested in the spatial variation of the 
static correlation function for $\delta |\Delta|_q$ at zero temperature, we 
evaluate the $T=0$ limit of 
$X_{\Delta}({\bf q},\Omega_m=0)\equiv X_\Delta({\bf q})$, 
\begin{eqnarray} 
\label{weakcorr}
X_{\Delta}({\bf q})& = &\left[\frac{1}{U}-\frac{1}{2}D({\bf q})\right]^{-1},\\
\label{dq}
D({\bf q})& = & \frac{1}{N_s}\sum_{\bf k} \frac{\gamma_{\bf k}^2}{E_+ + E_-}
\left(1+\frac{\xi_+\xi_- -\Delta_+\Delta_-}{E_+E_-}\right),
\end{eqnarray}
where, with a standard notation, $\Delta_{\bf k}=\Delta_0\gamma_{\bf k}$
\cite{notadelta}, $E_{\bf k}=\sqrt{\xi_{\bf k}^2+\Delta_{\bf k}^2}$, and 
$\xi_{\pm}$, $\Delta_{\pm}$, $E_{\pm}$ are calculated at momenta 
${\bf k} \pm {\bf q}/2$  respectively.

\begin{figure}[hbt]
\begin{center}
\includegraphics[width=8cm, angle=0]{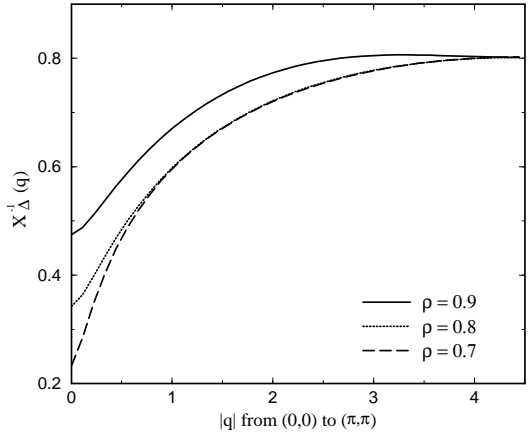}
\end{center}
\caption{ $X_{\Delta}^{-1}({\bf q})$ at intermediate coupling $U/t=1$ in the 
case of $d$-wave pairing for some values of the density $\rho$. The wave  
vector ${\bf q}$ varies along the diagonal of the Brillouin zone, and an 
analogous behavior is observed by varying ${\bf q}$ along different 
directions. As it is expected in the weak-coupling regime, 
$X_\Delta^{-1}({\bf q})$ exhibits a minimum at ${\bf q}=0$ [which corresponds 
to a maximum for the correlation function $X_\Delta({\bf q})$ in Eq. 
(\ref{weakcorr})].} 
\label{figweak}
\end{figure}

The length scale which controls the long-distance behavior of 
$X_{\Delta}({\bf r})$ may be extracted from the dominant part of its Fourier 
transform $X_\Delta({\bf q})$. It is generally expected that the main 
contribution to $X_\Delta({\bf q})$ comes from the region ${\bf q}\approx 0$. 
This is confirmed by our numerical calculations. In Fig. \ref{figweak} we 
show the ${\bf q}$ dependence of $X_\Delta^{-1}({\bf q})$ at various 
densities $\rho$, for $U/t=1$, in the case of $d$-wave pairing. Similar 
results are found in the $s$-wave case. Expanding Eq. (\ref{weakcorr}) for 
small momenta, we get \cite{notaxi}
\begin{equation}
X_{\Delta}({\bf q}) \simeq \frac{1}{m^2+c{\bf q}^2}, 
\label{weakesp}
\end{equation}
where the ``mass term'' is given by
\begin{equation}
m^2=\frac{1}{N_s}
\sum_{\bf k}  \frac{\gamma_{\bf k}^2\Delta_{\bf k}^2}{2E_{\bf k}^3},
\label{weakmass}
\end{equation}
and 
\begin{eqnarray} 
c&=&\frac{1}{16N_s}\sum_{\bf k}\left\{
\frac{2\gamma_{\bf k}^2}{E_{\bf k}^7}\xi_{\bf k}\Delta_{\bf k}
(2\Delta_{\bf k}^2-3\xi_{\bf k}^2)
({\partial_{\bf k}}\xi_{\bf k})\cdot({\partial_{\bf k}}
\Delta_{\bf k})\right.\nonumber\\
&+&\frac{\gamma_{\bf k}^2}{E_{\bf k}^5}\left[\xi_{\bf k}
(\xi_{\bf k}^2-2\Delta_{\bf k}^2)
\partial_{\bf k}^2\xi_{\bf k}+3\Delta_{\bf k}\xi_{\bf k}^2
\partial_{\bf k}^2\Delta_{\bf k}\right] \nonumber\\
&+&\left.\frac{\gamma_{\bf k}^2}{E_{\bf k}^7}\left[
(\xi_{\bf k}^4-4\xi_{\bf k}^2\Delta_{\bf k}^2)
({\partial_{\bf k}}\Delta_{\bf k})^2 +5\Delta_{\bf k}^2\xi_{\bf k}^2
({\partial_{\bf k}}\xi_{\bf k})^2\right]
\right\}.
\label{weakcoeff} 
\end{eqnarray}
When the relevant fluctuations are those near ${\bf q}=0$, $X_\Delta({\bf q})$ 
is maximum at ${\bf q}=0$, and $c>0$. In such a case the approximate 
expression (\ref{weakesp}) can be used to determine the long-distance 
behavior of $X_\Delta({\bf r})$. Indeed the Fourier transform of Eq. 
(\ref{weakesp}) has an exponential decay in real space with a characteristic 
length scale
\begin{equation}
\xi_0=\sqrt{\frac{c}{m^2}}.
\label{weakdef}
\end{equation}
As we show in Appendix A, differently from $\xi_{pair}$ given by Eq. 
(\ref{csipair}), such a length scale, evaluated from Eqs. (\ref{weakmass})
and (\ref{weakcoeff}), is {\em finite} also in the presence of nodal 
quasiparticles, making it a natural candidate for the characteristic length 
of $d$-wave superconductors. In the weak-coupling regime ($U/t\ll 1$), 
$\xi_0$, Eq. (\ref{weakdef}), can be expressed both in the $s$-wave and in 
the $d$-wave case by means of the average values of the Fermi velocity 
($v_F$) and of the gap ($\Delta_F$) on the Fermi surface,
\begin{equation}
\xi_0\simeq \frac{v_F}{\sqrt{12}\,\Delta_F}.
\label{eqcsi0}
\end{equation}
Here we define the average value of a function $h_{\bf k}$ on the Fermi
surface as 
\begin{equation}
h_F\equiv \sqrt{\langle h^2_{\bf k}\rangle_F}=
\sqrt{\frac{\sum_{\bf k} h^2_{\bf k}\delta(\xi_{\bf k})}
{\sum_{\bf k} \delta(\xi_{\bf k})}}.
\label{average}
\end{equation}
In the $s$-wave case ($\gamma_{\bf k}=1$) $\Delta_F=\Delta_0$, and $\xi_0$ 
coincides within a numerical factor of order one with the Cooper-pair size 
(see Appendix A). In the $d$-wave case $\Delta_F=\Delta_0\gamma^d_F$, so that 
the effective gap $\Delta_F$ which appears in Eq. (\ref{eqcsi0}) is smaller 
than the maximum value of the gap at the Fermi surface. In both cases, 
however, in the weak-coupling regime, the correlation function for 
$\delta|\Delta|_q$ exhibits an exponential decay over a length scale of order 
$v_F/\Delta_0$, while the different gap symmetries only introduce a numerical 
factor $\gamma_F$. It is then natural to assume such a length scale as the 
spatial cutoff for phase fluctuations in the effective phase-only action, 
both in $s$- and in $d$-wave superconductors.

\section{The strong-coupling regime}
The extension of the above results to the strong-coupling regime $U/t\gg 1$ 
is quite intriguing. Indeed, as the pairing increases the coefficient $c$ of 
Eq. (\ref{weakcoeff}) decreases, and becomes {\em negative}, making the 
definition (\ref{weakdef}) meaningless. On the other hand, it is commonly 
expected that in the diluted regime the fermionic system maps into a (weakly 
interacting) bosonic system, where the Cooper pairs act as boson particles, 
with a weak residual repulsion between them \cite{micnas}. It is well known 
\cite{fetter} that in a Bogoljubov liquid of weakly interacting bosons, the 
coherence length $\xi_{bos}$ which controls the correlations of the 
superfluid order parameter diverges as the density decreases. According to 
Ref. \cite{fetter}
\begin{equation}
\xi_{bos}=\frac{1}{\sqrt{2 g m_B  \rho_B}},
\label{csibos}
\end{equation}
where $g$ is the local repulsion, $m_B$ and $\rho_B$ are the mass and the 
particle density of bosons respectively. One would like to find a similar 
behavior for $\xi_0$ within the Hubbard model at strong coupling and in the 
low-density limit. As we anticipated in Sec. I, in the following we devote 
our attention to the case of $s$-wave pairing, which allows for an analytical 
treatment. However, one expects similar results also in the case of $d$-wave 
pairing. 

At strong coupling the definition (\ref{weakdef}) must be generalized to 
include two effects: (i) the fluctuations of $|\Delta|$ are strictly tied to 
the fluctuations of $\rho$, in such a way that both contribute to the 
correlation function for $|\Delta|$; when this interplay is considered, we 
find that (ii) the dominant contribution to $X_\Delta({\bf q})$ arises from 
${\bf q}\neq 0$ (except for the small-density regime), and the small-momentum 
approximation (\ref{weakesp}) is no longer appropriate. 

We first address the point (i). In the $s$-wave Hubbard model the interaction 
$H_I$ can be decoupled both in the particle-particle and in the particle-hole 
channel. When the density fluctuations are taken into account on the same 
footing as the Cooper-pair fluctuations, the effective action (\ref{effact}) 
gets modified, and reads
\begin{equation}
{\cal S}_{eff}=\sum_q\sum_{\mu=1,2\atop\nu=1,2} \Phi^\mu_q X_{\mu\nu}^{-1}(q)
\Phi^\nu_{-q},
\label{densact} 
\end{equation}
where $\Phi^{\nu=1}_q=\delta\rho_q$,$\Phi^{\nu=2}_q=\delta|\Delta|_q$, 
$X_{11}^{-1}(q)={1/U}-{1\over2}\chi_\rho(q)$,
$X_{12}^{-1}(q)=X_{21}^{-1}(q)=\chi_{\rho\Delta}(q)$, and 
$X_{22}^{-1}(q)={1/U}-{1\over2}D(q)$. Below, we only need the expressions in 
the static limit,
\begin{eqnarray}
\label{chiro}
\chi_\rho({\bf q})&=&\frac{1}{N_s}\sum_{\bf k} 
\frac{1}{E_+ + E_-}\left(1+\frac{\Delta_0^2-\xi_+\xi_-}{E_+E_-}\right),\\
\label{chird}
\chi_{\rho\Delta}({\bf q})&=& -\frac{\Delta_0}{2N_s}\sum_{\bf k} 
\frac{1}{E_+ + E_-}\;\frac{\xi_+ +  \xi_-}{E_+E_-}.
\end{eqnarray}

To take into account the effect of density fluctuations on $X_\Delta$ we 
integrate out the density-fluctuation field $\delta\rho_q$ in Eq. 
(\ref{densact}), and recover the action for $\delta|\Delta|$ only. The 
correlation function $X_\Delta$ now reads
\begin{equation}
X_\Delta(q)=\left[\frac{1}{U}-\frac{1}{2}D(q)-\frac{\chi_{\rho\Delta}^2(q)}
{\frac{1}{U}-\frac{1}{2}\chi_{\rho}(q)}\right]^{-1}.\label{rpa}
\end{equation}
It is worth noting that integrating out the density at the Gaussian level
corresponds to performing the RPA resummation for the correlation function
$X_\Delta$ in the particle-hole channel. At weak coupling the bubble
$\chi_{\rho\Delta}$ which couples the two channels is negligible, and the
result (\ref{weakcorr}) is recovered. In order to estimate Eq. (\ref{rpa}) at 
strong coupling, we evaluate the ${\bf q}$-dependent leading order in $t/U$
of the bubbles (\ref{dq}), (\ref{chiro}), and (\ref{chird}):
\begin{eqnarray}
\label{scdq}
D({\bf q})&\simeq& D(0)-c_\Delta w({\bf q})\nonumber\\
&\equiv&{1\over N_s}\sum_{\bf k} \frac{\xi_{\bf k}^2}{2 E_{\bf k}^3}
-c_\Delta w({\bf q}),\\
\label{scchird}
\chi_{\rho\Delta}({\bf q})&\simeq& \chi_{\rho\Delta}(0)- 
c_{\rho\Delta} w({\bf q})\nonumber\\
&\equiv&-{\Delta_0\over N_s}\sum_{\bf k} 
\frac{\xi_{\bf k}}{2 E_{\bf k}^3}- c_{\rho\Delta} w({\bf q}),\\
\label{scchi0}
\chi_\rho({\bf q})&\simeq& \chi_\rho(0)-c_\rho w({\bf q})\nonumber\\ 
&\equiv&{\Delta_0^2\over N_s}\sum_{\bf k} \frac{1}{2 E_{\bf k}^3}-c_\rho 
w({\bf q}),
\end{eqnarray}
where the function  
$w({\bf q})\equiv 4(\sin^2\frac{q_x}{2}+\sin^2\frac{q_y}{2})$ reduces to 
${\bf q}^2$ at small momenta, while respecting the lattice periodicity at 
higher momenta. In Eqs. (\ref{scdq})-(\ref{scchi0}) the ${\bf q}=0$ values 
are of order $1/U$, while the coefficients $c_\Delta, c_{\rho\Delta}$ and 
$c_\rho$ are of order $t^2/U^3$. All these coefficients are functions 
of the gap amplitude $\Delta_0$ and of the chemical potential $\mu$, whose 
dependence on the density can be determined by using the self-consistent 
saddle-point equations 
\begin{eqnarray*}
1&=&\frac{U}{2N_s}\sum_{\bf k}\frac{1}{E_{\bf k}},\nonumber\\
\delta&=&{1\over N_s}\sum_{\bf k} \frac{\xi_{\bf k}}{E_{\bf k}},\nonumber
\end{eqnarray*}
where we introduced the ``doping'' $\delta\equiv 1-\rho$ to simplify the 
notation \cite{notamu}. The above equations can be explicitly solved at 
strong coupling giving 
\begin{eqnarray*}
\Delta_0&=&\frac{U}{2}\left[(1-2\alpha^2)\sqrt{1-\delta^2}+
{\cal O}\left({\alpha^4}\right)\right],\\
\mu&=&-\frac{U}{2}\left[
\delta (1+4\alpha^2)+{\cal O}\left({\alpha^4}\right)\right],
\end{eqnarray*}
where we introduced the small parameter $\alpha\equiv2t/U$. 

\begin{figure}[hbt]
\begin{center}
\includegraphics[width=8cm, angle=0]{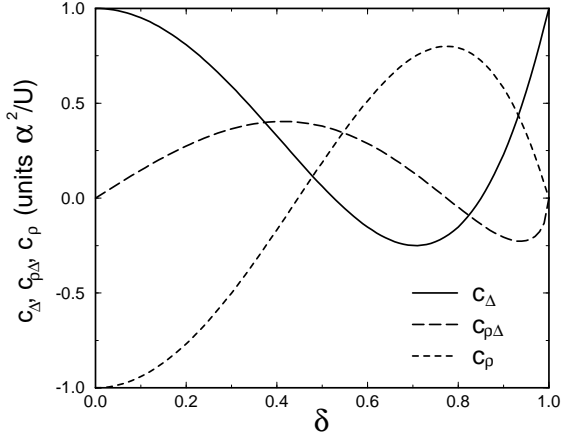}
\end{center}
\caption{ Doping dependence of the coefficients $c_\Delta, c_{\rho\Delta}$ and 
$c_\rho$ given by Eqs. (\ref{cd}), (\ref{crod}) and (\ref{cro}) 
respectively.}    
\label{figcoeff}
\end{figure}

As anticipated above, we shall investigate the region $0\leq \rho\leq 1$, i.e. 
$0\leq \delta\leq 1$, while the range $1\leq \rho\leq 2$ can be recovered by 
particle-hole symmetry ($\delta\to -\delta$). 

With long but straightforward calculations it can be seen that at leading 
order in $\alpha$ 
\begin{eqnarray}
\label{cd}
c_\Delta&=&\frac{\alpha^2}{U}(5\delta^4-5\delta^2+1),\\
\label{crod}
c_{\rho\Delta}&=&\frac{\alpha^2}{2U}\delta(3-5\delta^2)\sqrt{1-\delta^2},\\
\label{cro}
c_\rho&=&\frac{\alpha^2}{U}(1-\delta^2)(5\delta^2-1),
\end{eqnarray}
whose behavior is reported in Fig. \ref{figcoeff}.

By means of Eqs. (\ref{scdq})-(\ref{scchi0}) and Eqs. (\ref{cd})-(\ref{cro}) 
we obtain the ${\bf q}$-dependent strong-coupling expression of $X_\Delta$ at 
leading order in $\alpha$,
\begin{eqnarray}
X_\Delta^{-1}({\bf q})&=&
\frac{1}{U}-\frac{1}{2}\left[D(0)-c_\Delta w({\bf q})\right]\nonumber\\
&-&\frac{\chi^2_{\rho\Delta}(0)}
{\frac{1}{U}-\frac{1}{2}\left[\chi_\rho(0)-c_\rho w({\bf q})\right]}\nonumber\\
&+&\frac{2\chi_{\rho\Delta}(0)c_{\rho\Delta}w({\bf q}) }
{\frac{1}{U}-\frac{1}{2}\left[\chi_\rho(0)-c_\rho w({\bf q})\right]},
\label{sc}
\end{eqnarray}
where we neglect the subleading $c_{\rho\Delta}^2$ term. 

\subsection{Low-density regime}
In the limit of low 
density $(\delta\approx 1)$ it is possible to rewrite Eq. (\ref{sc}) in a 
simpler form. Indeed, at the leading order in $\alpha$ the denominator of the 
third and fourth terms in Eq. (\ref{sc}) is given by
$$
\frac{1}{U}-\frac{1}{2}\left[\chi_\rho(0)-c_\rho w({\bf q})\right]
\simeq \frac{1}{U}\left[\delta^2+4\alpha^2-{\alpha^2\over 2}w({\bf q})\right].
$$
As a consequence, for $\alpha \ll \delta$ we can put 
\begin{eqnarray}
\frac{1}{\frac{1}{U}-\frac{1}{2}\left[\chi_\rho(0)-c_\rho w({\bf q})\right]}
\nonumber\\
\approx\frac{1}{\frac{1}{U}-\frac{1}{2}\chi_\rho(0)}
\left[1-\frac{\frac{1}{2}c_\rho w({\bf q})}
{\frac{1}{U}-\frac{1}{2}\chi_\rho(0)}\right].\label{greatd}
\end{eqnarray}
Substituting Eq. (\ref{greatd}) into Eq. (\ref{sc}) we have
\begin{equation}
X_{\Delta}({\bf q})=\frac{1}{M^2+C w({\bf q})},
\label{strongesp}
\end{equation}
where the ${\bf q}=0$ value $M^2\equiv X^{-1}_\Delta(0)$ is given by
\begin{eqnarray}
\label{0mass}
M^2=\frac{1-\delta^2}{U}\left[1-6\alpha^2+30\delta^2\alpha^2\right.\nonumber\\
-\left. \frac{\delta^2(1-36\alpha^2+60\alpha^2\delta^2) }{\delta^2+4\alpha^2
(1-\frac{15}{2}\delta^2+\frac{15}{2}\delta^4)} \right] 
\tende{\alpha\ll\delta}\frac{4\alpha^2}{U}\frac{1-\delta^2}{\delta^2},
\end{eqnarray}
and
\begin{eqnarray}
\label{0coeff}
C &=& \frac{\alpha^2}{4U}\left[ 
1-5\delta^2+5\delta^4+\frac{2\delta^2(1-\delta^2)(5\delta^2-3)}
{\delta^2+4\alpha^2}+\right.\nonumber\\
&+& \left.\frac{\delta^2(5\delta^2-1)(1-\delta^2)^2}
{\left(\delta^2+4\alpha^2\right)^2}\right]\tende{\alpha\ll\delta} 
\frac{2\alpha^2}{U}\frac{2\delta^2-1}{\delta^2}.
\end{eqnarray}

Therefore, $C>0$ in the diluted limit ($\delta\approx 1$), and 
$X_\Delta({\bf q})$ is dominated by the small-momentum region, where 
$w({\bf q})\approx{\bf q}^2$ and the coherence length is given by the 
generalization of Eq. (\ref{weakdef}), with $c$ and $m^2$ substituted by $C$ 
and $M^2$ respectively,
\begin{equation}
\xi_0=\sqrt{\frac{C}{M^2}}=\sqrt{\frac{2\delta^2-1}{8(1-\delta^2)}}
\tende{\rho \rightarrow 0\atop{(\delta \rightarrow 1)}} \frac{1}{4\sqrt \rho}.
\label{dil}
\end{equation}
The result (\ref{dil}) for the coherence length $\xi_0$ shows, at low particle 
density, the same divergence of the bosonic coherence length $\xi_{bos}$ 
given by Eq. (\ref{csibos}). Notice that a more strict comparison  between Eq.
(\ref{dil}) and Eq. (\ref{csibos}) requires a dependence of $\xi_0$ on the
effective mass of the electron pair $m_P$ and on the pair-pair residual 
repulsion $g_P$ in the bosonic limit, of the form 
$\xi_0^2\simeq 1/(g_P m_P \rho)$. This is indeed the case, since in the 
bosonic limit of the fermionic model $\rho_B=\rho/2$, $m_P\simeq U/8t^2$, and 
the residual repulsion of the bosonic model corresponds to the inverse of the 
compressibility of the fermionic system $g_P\sim \chi^{-1}$. Following the 
analysis of Ref. \cite{noi2}, it can be seen that at strong coupling the 
compressibility of the Hubbard model is $\chi= U/8t^2$ \cite{noiunpu}, 
leading to the explicit cancellation of $t$ and $U$ in Eq. (\ref{csibos}), in
agreement with Eq. (\ref{dil}). This supports the reliability of the mapping 
of the diluted large-$U$ Hubbard model into an effective boson model 
\cite{micnas,depalo}.

\subsection{High-density regime}
\begin{figure}[hbt]
\begin{center}
\includegraphics[width=8cm, angle=0]{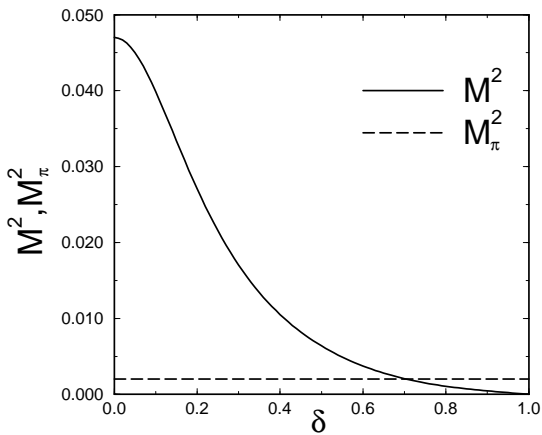}
\end{center}
\caption{ Doping dependence of $M^2$ and $M_\pi^2$ given by Eqs. 
(\ref{0mass}) and (\ref{pimass}) respectively, for $t=1$ and $U=20$.}    
\label{figmass}
\end{figure}

As $\delta$ decreases, according to Eq. (\ref{0coeff}), the coefficient $C$ 
decreases and becomes {\em negative} for $\delta<1/\sqrt 2$. This change of 
sign is not due to a failure of our approximation, since at strong coupling 
$\alpha\ll 1$, so that at $\delta=1/\sqrt 2$ the simplified expression 
(\ref{strongesp}) is still valid. The vanishing of the leading coefficient in 
the small-${\bf q}$ expansion is instead a signal of the fact that the 
maximum of the correlation function $X_\Delta({\bf q})$ is moving to a finite 
${\bf q}$. Since the ${\bf q}$-dependence of $X_\Delta({\bf q})$ comes 
entirely from the function $w({\bf q})$, the only candidate alternative to 
${\bf q}=(0,0)$ is ${\bf q}={\bf Q}\equiv (\pi,\pi)$. Indeed, evaluating  
$M^2_\pi\equiv X^{-1}_\Delta({\bf Q})$ according to Eq. (\ref{sc}) we find 
that
\begin{equation}
M^2_\pi=\frac{4\alpha^2}{U}
\label{pimass}
\end{equation}
at all doping, so that $M^2_\pi<M^2$ as far as $\delta<1/\sqrt{2}$, as 
depicted in Fig. \ref{figmass}. In such a case, the dominant 
$\delta|\Delta|_q$ modes are located near the wave vector ${\bf Q}$. This is
an effect of the coupling of the Cooperon to the density, which is a massless 
mode at exactly half filling ($\delta=0$) for ${\bf q}={\bf Q}$ \cite{cdw}. 
Notice that this behavior is peculiar of the case $\rho=1$ for a band 
dispersion arising from a nearest-neighbor hopping. In the presence of an 
attractive on-site interaction the system exhibits an enlarged symmetry with 
respect to the instability in the particle-particle channel (at ${\bf q}=0$) 
or in the particle-hole channel (at ${\bf q}={\bf Q}$). Once the symmetry has 
been explicitly broken in the Cooper channel, the density becomes a Goldstone 
mode, similarly to the phase, reflecting such a degeneracy. Since at strong 
coupling $\delta\rho$ tends to fluctuate coherently with $\delta|\Delta|$ 
\cite{micnas,depalo}, at small doping and strong coupling both 
$X_\Delta({\bf q})$ and $X_\rho({\bf q})$ have a maximum at the wave vector 
which controls the instability of the density mode approaching half filling, 
i.e., ${\bf q}={\bf Q}$. 

As a result, for $M_\pi^2<M^2$ the long-distance behavior of 
$X_\Delta({\bf r})$ should be controlled by the characteristic length 
$\xi_\pi$ obtained by considering the expansion of $X_\Delta({\bf q})$ around 
${\bf Q}$, 
\begin{eqnarray}
X_\Delta({\bf r})&\simeq& \int d{\bf q} \frac{e^{i{\bf q} \cdot {\bf r}}}
{M^2_\pi+A({\bf q}-{\bf Q})^2}\nonumber\\
&=& e^{i{\bf Q}\cdot{\bf r}} \int d{\bf q}\frac{e^{i{\bf q} \cdot {\bf r}}}
{M^2_\pi+A{\bf q}^2},\label{trasf}
\end{eqnarray}
so that the resulting $X_\Delta({\bf r})$ is a staggered function with an 
exponential envelope controlled by the stiffness $A$ of the ${\bf q}$-modes 
near ${\bf Q}$ and by the mass $M_\pi^2$. The parameter $A$ is obtained by 
evaluating, at ${\bf q}={\bf Q}$, the second-order derivative of 
$X_\Delta^{-1}({\bf q})$, as given by the expression (\ref{sc}),
\begin{equation}
A\equiv\left.\frac{1}{4}\partial_{\bf q}^2 X^{-1}_\Delta({\bf q})
\right\vert_{{\bf q}={\bf Q}} =\frac{\alpha^2}{2U}\frac{1-2\delta^2}{\delta^2}.
\label{picoeff}
\end{equation}
According to Eq. (\ref{trasf}), the long-distance decay of $X_\Delta({\bf r})$
is exponential, with a characteristic length scale 
\begin{equation}
\xi_\pi=\sqrt{\frac{A}{M_\pi^2}}=\sqrt{\frac{1-2\delta^2}{8\delta^2}},
\label{csipi}
\end{equation}
which matches continuously at $\delta=1/\sqrt 2$ with the $\xi_0$ 
previously defined in Eq. (\ref{dil}).

\begin{figure}[hbt]
\begin{center}
\includegraphics[width=8cm, angle=0]{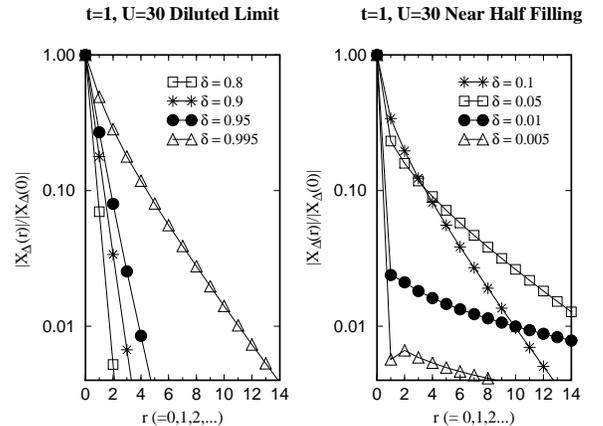}
\end{center}
\caption{Absolute value of $X_\Delta({\bf r})$ normalized to its ${\bf r}=0$ 
value for ${\bf r}=n {\bf \hat x}$ at various $\delta$. Left panel: the slope 
of the long-distance (exponential) decay of $|X_\Delta({\bf r})|$ in the 
diluted limit is given by $-\xi_0^{-1}$, with $\xi_0$ from Eq. (\ref{dil}). 
Right panel: $|X_\Delta({\bf r})|$ near half filling. Although the 
characteristic length $\xi_\pi$ of the exponential tail increases as the 
doping decreases, $|X_\Delta({\bf r})|$ is strongly suppressed at much shorter 
distances.}
\label{figtrasf}
\end{figure}

According to our strong-coupling expression (\ref{sc}) both $\xi_0$ and 
$\xi_\pi$ vanish at $\delta=1/\sqrt{2}$. This is clearly an artifact of 
retaining only the order ${\cal O}(\alpha^2)$ in the ${\bf q}$ dependence of 
$X_\Delta^{-1}({\bf q})$. By including in Eqs. (\ref{scdq})-(\ref{scchi0}) 
the next terms, which are of the form 
$\sim\alpha^4\{w^2({\bf q})+{\rm const}\times[\sin^4(q_x/2)+\sin^4(q_y/2)]\}$, 
the ratio between the coefficient of ${\bf q}^4$ and the mass term in Eq. 
(\ref{strongesp}) gives a coherence length of order 
$\xi_0\sim\xi_\pi\sim{\cal O}(\alpha^2)$.

\begin{figure}[hbt]
\begin{center}
\includegraphics[width=8cm, angle=0]{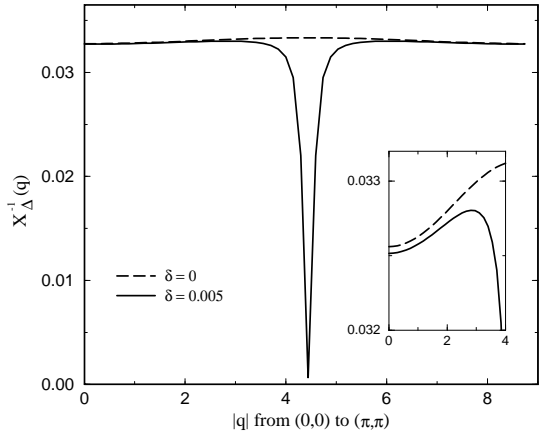}
\end{center}
\caption{$X_\Delta^{-1}({\bf q})$ at $t=1, U=30$ for $\delta=0.005<4\alpha^2$
(solid line) and $\delta=0$ (dashed line). ${\bf q}$ varies along the 
diagonal of the Brillouin zone. The width of the minimum at ${\bf Q}$ 
decreases as the doping decreases, vanishing at half filling. Inset: detail of 
the behavior of $X_\Delta^{-1}({\bf q})$ away from ${\bf Q}$.}
\label{figddq}
\end{figure}

One usually defines the correlation length $\xi_0$ as 
$$
\xi_0^{-1}= -\lim_{r\rightarrow \infty} \frac{\ln X_\Delta(r)}{r}.
$$
According to this definition, the identification between $\xi_0$ the and 
$\xi_\pi$ for $\delta<1/\sqrt 2$ would imply a divergent $\xi_0$ at 
half filling. Such a divergence is surprising, since in this regime one would 
naively expect that the correlation length for the amplitude fluctuations 
attains a value of the order of the lattice constant. Indeed this problem is 
only apparent, as we can see by evaluating numerically  $X_\Delta({\bf r})$
near half filling from Eqs. (\ref{chiro})-(\ref{rpa}). 
In Fig. \ref{figtrasf} we report $X_\Delta({\bf r})$ normalized to its 
${\bf r}=0$ value at various doping, both in the diluted (left panel) and 
nearly half-filled (right panel) limit. In particular, near half filling, 
it can be seen that, although the long-distance behavior of 
$X_\Delta({\bf r})$ is controlled by the large length scale $\xi_\pi$, 
Eq. (\ref{csipi}), which effectively diverges as $\delta\rightarrow 0$, 
$X_\Delta({\bf r})$ is severely suppressed with respect to its ${\bf r}=0$ 
value at much shorter distances, of the order of the lattice spacing. A 
simple criterion to estimate the coherence length consists in determining
the distance $\xi_\kappa$ at which $X_\Delta({\bf r})$ has reached a fixed 
percent $\kappa$ of its ${\bf r}=0$ value, say $\kappa=0.1$, in Fig. 
\ref{figtrasf}. One can identify $\xi_0$ as the minimum value between 
$\xi_\kappa$ and the lattice spacing, which is the lower bound for $\xi_0$
in a lattice model. From Fig. \ref{figtrasf} it follows that the maximum 
value of $\xi_0$ at $U/t=30$ is about $5$ lattice spacings, and is reached
at $\delta=0.05$. At lower doping the decay of $X_\Delta({\bf r})$ is much 
more rapid, even though the long-distance tail extends over a greater 
distance. On the other hand, in the low-density regime the coherence length 
is really diverging. In the left panel of Fig. \ref{figtrasf} we report 
$|X_\Delta({\bf r})|/|X_\Delta(0)|$ at high doping. $X_\Delta({\bf r})$ is 
exponentially decreasing in this regime, with the characteristic length
given by Eq. (\ref{dil}). Indeed, as $\delta$ increases, the slope of the 
curve, i.e.  $-\xi_0^{-1}$, decreases, and the {\em overall} decay of 
$X_\Delta({\bf r})$ extends up to larger distances.

The different role played by $\xi_0$, Eq. (\ref{dil}), and $\xi_\pi$, Eq. 
(\ref{csipi}), in determining the decay of $X_\Delta({\bf r})$ can be better 
understood by analyzing in greater detail the ${\bf q}$ dependence of 
$X_\Delta({\bf q})$ approaching half filling. $X_\Delta^{-1}({\bf q})$ is 
reported in Fig. \ref{figddq}. Let us consider the curve for $\delta=0.005$. 
Even though the minimum of $X_\Delta^{-1}$ is at ${\bf q}={\bf Q}$, the width 
of this minimum decreases with $\delta$. This reflects the fact that, 
according to Eq. (\ref{picoeff}) and (\ref{pimass}), the divergence of 
$\xi_\pi$, Eq. (\ref{csipi}) at half filling arises from an increasing 
stiffness $A$ of the ${\bf q}$ modes near ${\bf Q}$ with a fixed mass. As a 
consequence near half filling $X_\Delta({\bf q})$  is almost constant (and 
equal to its ${\bf q}=0$ value) except in a narrow region near ${\bf Q}$, 
which then contributes to the integral (\ref{trasf}) only at very large 
distances and with a small weight.

In the inset of Fig. \ref{figddq} it is shown that at $\delta=0.005$ a local 
minimum of $X_\Delta^{-1}({\bf q})$ at ${\bf q}=0$ exists besides the global 
minimum at ${\bf q}={\bf Q}$. Indeed, the second-order derivative of 
$X_\Delta^{-1}$ at ${\bf q}=0$, given by $C$ of Eq. (\ref{0coeff}) is again 
positive at $\delta\leq 4 \alpha^2$. At exactly $\delta=0$ only the 
${\bf q}=0$ minimum survives, since, according to Eqs. 
(\ref{chird}) and (\ref{sc}), $\chi_{\rho\Delta}=0$ so that the amplitude 
fluctuations decouple from the density fluctuations and $X_\Delta({\bf q})$ 
reduces to the form (\ref{weakcorr}). As a consequence, the maximum of 
$X_\Delta({\bf q})$ is again at zero momentum and the coherence length is
$$
\xi_0(\delta=0)=
\sqrt{\frac{c_\Delta}{{2\over U}-D(0)}}=\sqrt{\frac{\alpha^2}{2}}.
$$
This small value matches continuously with the characteristic length scale 
which controls the short-distance decay of $X_\Delta({\bf r})$ 
near half filling, and can be extracted form the data reported in the right 
panel of Fig. \ref{figtrasf}. 

\subsection{The density-density correlation function}
In the last part of this section, to gain some more insight into the physics 
of the interplay between the particle-particle and particle-hole channels at 
strong coupling, we compare the above results for the pairing correlation 
function with the outcomes of the same calculations for the density-density 
correlation function. In analogy with the case of $X_\Delta(q)$,
starting from the same Eq. (\ref{densact}), we find that the density-density 
correlation function, defined as
$X_{\rho}({\bf q},\Omega_m)=2<\delta\rho_q \delta\rho_{-q}>$,
is given by
\begin{equation}
X_\rho(q)=\left[\frac{1}{U}-\frac{1}{2}\chi_\rho(q)-
\frac{\chi_{\rho\Delta}^2(q)}
{\frac{1}{U}-\frac{1}{2}D(q)}\right]^{-1},
\label{rpaden}
\end{equation}
i.e. the RPA resummation for $X_\rho$ in the particle-particle channel. 
Evaluating Eq. (\ref{rpaden}) in the strong-coupling regime by means of
Eqs. (\ref{scdq})-(\ref{scchi0}) and Eqs. (\ref{cd})-(\ref{cro}), we find
that at leading order in $\alpha$ 
$X^{-1}_\rho({\bf q})={\cal M}^2+{\cal C} w({\bf q})$,
where ${\cal M}^2\equiv X^{-1}_\rho(0)=4\alpha^2/U$, and
${\cal C}=(\alpha^2/2U)(2\delta^2-1)/(1-\delta^2)$. It follows that 
${\cal C}>0$ at $\delta>1/\sqrt{2}$, so that the maximum of 
$X_\rho({\bf q})$ is at ${\bf q}=0$ and the coherence length $\xi_\rho$, 
which controls the exponential decay of $X_\rho({\bf r})$, does coincide with 
$\xi_0$, Eq. (\ref{dil}),
\begin{equation}
\xi_{\rho}=\sqrt{\frac{\cal C}{{\cal M}^2}}=
\sqrt{\frac{2\delta^2-1}{8(1-\delta^2)}}.\label{csirold}
\end{equation}
Instead, for $\delta<1/\sqrt{2}$, ${\cal C}<0$ and the maximum of 
$X_\rho({\bf q})$ is at ${\bf q}={\bf Q}$. Indeed, the mass term  for the
density-density correlation function at ${\bf q}={\bf Q}$ is
\begin{equation}
{\cal M}_\pi^2\equiv X_\rho^{-1}({\bf Q})={\cal M}^2+8{\cal C}=
\frac{4\alpha^2}{U}\,\frac{\delta^2}{1-\delta^2},
\label{mropi}
\end{equation}
and the second-order derivative at ${\bf q}={\bf Q}$ is
$$
{\cal A}\equiv\left.\frac{1}{4}\partial_{\bf q}^2 X^{-1}_\rho({\bf q})
\right\vert_{{\bf q}={\bf Q}} =-{\cal C}.
$$
As a consequence, for $\delta<1/\sqrt{2}$, ${\cal M}_\pi^2<{\cal M}^2$, and 
the function $X_\rho({\bf r})$ oscillates with an exponential envelope 
controlled by the characteristic length
\begin{equation}
\xi_\rho=\sqrt{\frac{\cal A}{{\cal M}^2_\pi}}=
\sqrt{\frac{1-2\delta^2}{8\delta^2}},
\label{csirohd}
\end{equation}
which coincides with $\xi_\pi$, Eq. (\ref{csipi}). However, contrary to the
case of the $|\Delta|$ mode, which is suppressed at much shorter distances
due to the presence of a second small length scale associated with the
secondary minimum of $X_\Delta^{-1}({\bf q})$ at ${\bf q}=0$, the exponential 
decay of $|X_\rho({\bf r})|$ is controlled by a single length scale given by
Eq. (\ref{csirohd}), which diverges at $\delta=0$. This divergency is due to 
the fact that the density mode becomes a Goldstone mode for the enlarged 
symmetry of the Hubbard model at half filling, so that the mass 
${\cal M}_\pi^2$, Eq. (\ref{mropi}), vanishes, while the stiffness ${\cal A}$ 
is finite. As a consequence, while $\xi_0$ for pairing fluctuations is 
ultimately cut off by the lattice constant, the divergence of $\xi_\rho$ as 
$\xi_\pi$ according to Eq. (\ref{csirohd}) is the signal of the degenerate 
instability at $\rho=1$ \cite{cdw}. On the other hand, since in the diluted 
limit the pairing and density fluctuations become proportional, as shown in 
Ref. \cite{depalo}, the characteristic length of the spatial decay of both 
pairing [Eq. (\ref{dil})] and density mode [Eq. (\ref{csirold})] diverges 
according to the mapping with the bosonic model. We like to point out that 
the degenerate instability at $\rho=1$ is specific of the Hubbard model with 
nearest-neighbor hopping. Indeed, as discussed in Appendix B, the inclusion 
of a next-to-nearest-neighbor hopping term $t'$ spoils this degeneracy and 
the charge sector becomes massive even at $\rho=1$. However, by approaching 
$\rho=1$, the charge mode at ${\bf q}={\bf Q}$ still affects the pairing 
correlation function as discussed in this section, provided $t'/t$ is not too 
large.
 
The extension of the above results to the strong-coupling regime for the
$d$-wave symmetry is not straightforward. Indeed, the decoupling of the
interaction $H_I$, Eq. (\ref{mod}), in the particle-hole channel does not
simply reduce to a density-density interaction, but rather renormalizes the
bare band dispersion, limiting the possibility for an analytical treatment.
Nevertheless, supported also by the analysis of the weak-coupling case, we 
expect that in the strong-coupling limit results similar to those of the 
$s$-wave superconductor hold. In particular, one would find a coherence 
length of the order of the lattice parameter for all doping except in the 
low-density regime, where the mapping to the bosonic system should be 
recovered. Anyway, further numerical work is required to address this issue 
in more detail. 

\section{Conclusions}
In summary, we studied the evolution of the superconducting coherence length 
$\xi_0$ from weak to strong coupling. For both $s$-wave and $d$-wave 
superconductors, $\xi_0$ is defined through the spatial decay of the
correlation function $X_\Delta$ for the fluctuations of the modulus of the 
order parameter. At weak coupling the issue arises of the comparison between 
$\xi_0$ and the Cooper-pair size $\xi_{pair}$. These two length scales are 
both of order $v_F/\Delta_0$ in $s$-wave superconductors. However, as we 
showed, this identification of $\xi_0$ and $\xi_{pair}$ at weak coupling does 
not hold for $d$-wave superconductor, where $\xi_{pair}$ is divergent, while 
$\xi_0$ is finite and of order $v_F/\Delta_0$.

At strong coupling the modulus and density fluctuations are coupled, leading 
to a contribution of the density mode to the correlation function $X_\Delta$. 
The case of $s$-wave pairing within the negative-$U$ Hubbard model is simpler 
and allows for a detailed analytical treatment. We then evaluated $\xi_0$ for 
$s$-wave superconductors by properly including density fluctuations. We found 
that the coherence length in the diluted regime diverges as 
$\xi_0\sim 1/\sqrt \rho$, according to the mapping of the Hubbard model into 
a weakly-interacting effective bosonic model, in the strong-coupling and 
low-density regime. Away from the diluted limit the coherence length attains 
a value of order of few lattice spacings, as reflecting the local character 
of the Cooper pairing. Similar results are expected in the case of $d$-wave 
pairing, where however the analytical treatment cannot be carried out, and 
a deeper insight can only be gained by further numerical investigation.  

Finally, we comment on the issue of choosing an appropriate cutoff for the 
phase-only action when analyzing phase-fluctuation effects in superconductors. 
Our results indicate that at weak coupling it is reasonable to consider 
momenta up to a cutoff $|{\bf q}_c|\simeq \Delta_0/v_F$ for both $s$-wave 
and $d$-wave superconductors. In the strong-coupling regime, as we explicitly 
showed in the $s$-wave case, one should take $|{\bf q}_c| \simeq 1/a$ over a 
wide doping region, with the noticeable exception of the low-density region, 
where $\xi_0$ diverges and $|{\bf q}_c| \simeq \sqrt{\rho}/a$.
\vskip 0.5truecm
\par\noindent
{\bf Acknowledgments}: We thank M. Capone, C. Di Castro, M. Grilli, P. Pieri, 
and G. Strinati for stimulating discussions and suggestions.

\appendix
\section{Cooper-pair size and Coherence length at weak coupling}

In this appendix we report the detailed calculation of $\xi_{pair}$ and
$\xi_0$ in the $s$-wave and $d$-wave case at weak coupling. Let us start 
with $\xi_{pair}$, which is defined in Eq. (\ref{csipair}) as 
$\xi_{pair}=\sqrt{{\cal N}/{\cal D}}$ 
where
\begin{eqnarray}
{\cal N}&\equiv&\int d {\bf r}\, |\psi({\bf r})|^2 {\bf r}^2=
\frac{1}{N_s}\sum_{\bf k} [\partial_{\bf k}\phi({\bf k})]^2\nonumber\\ 
&=&\frac{1}{N_s}\sum_{\bf k}\left\{
\frac{\xi_{\bf k}^4}{E_{\bf k}^6}(\partial_{\bf k} \Delta_{\bf k})^2
+\frac{\Delta_{\bf k}^2\xi_{\bf k}^2}{E_{\bf k}^6} 
(\partial_{\bf k}\xi_{\bf k})^2\right.\nonumber\\
&-& \left.\frac{2\Delta_{\bf k}\xi_{\bf k}^3}{E_{\bf k}^6}
(\partial_{\bf k}\Delta_{\bf k}) \cdot (\partial_{\bf k}\xi_{\bf k})\right\}, 
\label{eqa}
\end{eqnarray}
and
$$
{\cal D}\equiv\int d {\bf r} \,|\psi({\bf r})|^2=
\frac{1}{N_s}\sum_{\bf k} \phi^2({\bf k})=
\frac{1}{N_s}\sum_{\bf k} \frac{\Delta^2_{\bf k}}{E^2_{\bf k}}.
$$
In the $s$-wave case only the second term in the right-hand side of Eq. 
(\ref{eqa}) survives, and letting 
${\bf v}_{\bf k}^2\equiv (\partial_{{\bf k}}\xi_{{\bf k}})^2$ we have
\begin{eqnarray*}
{\cal N} &=& \frac{1}{N_s}\sum_{{\bf k}} \frac{\Delta_0^2\xi_{{\bf k}}^2 
{\bf v}_{{\bf k}}^2}{E_{{\bf k}}^6}\nonumber\\
&=& \Delta_0^2 \int dx  \frac{x^2}{(x^2+\Delta_0^2)^3} 
\frac{1}{N_s}\sum_{{\bf k}} v^2_{{\bf k}} \delta(x-\xi_{{\bf k}})\nonumber\\
&=&   \Delta_0^2 \int dx  \frac{x^2}{(x^2+\Delta_0^2)^3} V(x),
\end{eqnarray*}
where we define 
$V(x)\equiv(1/N_s)\sum_{\bf k}{\bf v}^2_{\bf k}\delta(x-\xi_{\bf k})$. 
Since at weak coupling the main contribution to the above integral comes 
from $x\simeq 0$, we take 
$V(x)\simeq V(0)=(1/N_s)\sum_{\bf k}{\bf v}^2_{\bf k}\delta(\xi_{\bf k})$ and
we extend the integral over $x$ between $-\infty$ and $+\infty$ to extract 
the leading behavior
$$
{\cal N} \simeq \Delta_0^2 V(0) \int_{-\infty}^{+\infty} 
dx \frac{x^2}{(x^2+\Delta_0^2)^3}=\frac{\pi V(0)}{8\Delta_0}. 
$$
Similarly,
$$
{\cal D} \equiv \frac{1}{N_s}
\sum_{\bf k} \frac{\Delta_0^2}{E_{\bf k}^2} = \Delta_0^2 \int dx 
\frac{N(x)}{x^2+\Delta_0^2} \simeq  \pi N(0)\Delta_0,
$$
where $N(x)\equiv(1/N_s)\sum_{\bf k}\delta(x-\xi_{{\bf k}})$ is the 
density of states in the metallic phase. Then we find
\begin{equation}
\xi_{pair}^2= \frac{v_F^2}{8\Delta_0^2}
\label{csis}
\end{equation}
where, according to Eq. (\ref{average}), we introduce the mean value of the
velocity at the Fermi surface
$$
v_F^2=<{\bf v}^2_{\bf k}>_F= \frac{V(0)}{N(0)}.
$$
It is worth noting that the approximations leading to Eq. (\ref{csis}) are
no longer valid when the chemical potential approaches the extrema or the 
saddle points of the spectrum $\xi_{\bf k}$ (critical points), where a more 
refined evaluation 
is needed to get the correct result. However, by means of numerical 
calculations we checked that the estimate (\ref{csis}) is very accurate even 
close to the critical points.

In the $d$-wave case $\xi_{pair}$ is infinite, since the nodal quasiparticles
give a logarithmically divergent contribution to ${\cal N}$ in Eq. 
(\ref{eqa}). Near the nodes the quasiparticle spectrum $E_{\bf k}$ is 
cone-like, $\xi_{\bf k}\simeq v_{1} k_1$, $\Delta_{\bf k} \simeq v_{2} k_2$,
and $E_{\bf k}\simeq \sqrt{v_1^2 k_1^2+ v_2^2 k^2_2}$. Here $v_1$ and $v_2$ 
are the Fermi velocity and the slope of the gap $\Delta_{\bf k}$ at the node,
and $k_1,~k_2$ are the components of ${\bf k}$ along the directions 
perpendicular and parallel to the Fermi surface respectively, measured from 
the node. By introducing the polar coordinates $(E,\theta)$, such that when 
${\bf k}$ is near a node $\xi_{\bf k}= E \cos{\theta}$ and 
$\Delta_{{\bf k}}=E \sin{\theta}$, and using the identity,
$$
\frac{1}{N_s}\sum_{k_1 k_2}= \int\frac{ d \theta d E \, E}{ v_1 v_{2}},
$$
we find that the first two (positive) terms in the right-hand side of Eq. 
(\ref{eqa}) are divergent, while the last term, which in principle has not 
a definite sign, vanishes due to the orthogonality of 
$\partial_{\bf k}\Delta_{\bf k}$ and  $\partial_{\bf k}\xi_{\bf k}$ for a 
cone-like spectrum. For example, the first term gives
\begin{eqnarray*} 
&~&\frac{1}{N_s} \sum_{\bf k} \frac{\xi_{{\bf k}}^4}{E_{\bf k}^6} 
(\partial_{\bf k}\Delta_{\bf k})^2 \nonumber\\
&\simeq&\frac{v_1}{v_{2}}\int_{0}^{2\pi} d \theta \int_0^{\Lambda} d E 
\frac{E^5 \cos^4{\theta}}{E^6} \propto \int_0^{\Lambda}\frac{d E}{E}
 \rightarrow \infty,
\end{eqnarray*} 
where $\Lambda$ is a properly defined upper cutoff. Notice, instead, that the 
contribution of the nodal quasiparticles to ${\cal D}$ is finite,  
\begin{eqnarray*} 
{\cal D}&=&\frac{1}{N_s}\sum_{\bf k}\frac{\Delta^2_{\bf k}}{E^2_{\bf k}}
\nonumber\\
&\simeq&\frac{1}{v_1 v_{2}} \int_{0}^{2\pi} d \theta \int_0^{\Lambda} d E 
\frac{E^3 \sin^2{\theta}}{E^2} \propto \int_0^{\Lambda}d E E.
\end{eqnarray*} 

We next turn to the evaluation of the coherence length $\xi_0$ in the 
weak-coupling regime, defined in Eq. (\ref{weakdef}). It can be seen that at
weak coupling  the leading contribution to $c$ comes from the last term in
Eq. (\ref{weakcoeff}), so in Eq. (\ref{weakdef}) we must take
$$
c\simeq \frac{5}{16N_s}\sum_{\bf k} \frac{\gamma_{\bf k}^2}{E_{\bf k}^7}
\Delta_{\bf k}^2\xi_{\bf k}^2 {\bf v}_{\bf k}^2,
$$
while $m^2$ is given by Eq. (\ref{weakmass}). Observe that both $c$ and $m^2$
are finite whatever is the gap symmetry. In the $d$-wave case it can be seen 
that, thanks to the presence of the $\gamma^2_{\bf k}$ factor in the previous 
equation, the contribution of nodal quasiparticles to $c$ is indeed 
proportional to $\int d\theta dE E^7 \sin^4\theta\cos^2\theta/E^7$. Let us 
more generally consider the case of arbitrary gap modulation $\gamma_{\bf  k}$ 
and band dispersion $\xi_{\bf k}$. We first evaluate the mass term
$$
m^2= \frac{\Delta_0^2}{2N_s} \sum_{{\bf k}} \frac{\gamma_{\bf k}^4}
{(\xi_{\bf k}^2+\gamma_{\bf k}^2\Delta_0^2)^{\frac{3}{2}}}= 
\frac{\Delta_0^2}{2} \int d z \; 
\frac{G(z)}{(z^2+\Delta_0^2)^{\frac{3}{2}}},
$$
where $G(z)=(1/N_s)\sum_{\bf k}|\gamma_{\bf k}|
\delta\left(z -\xi_{\bf k}/\gamma_{\bf k}\right)$. Again, the main 
contribution to the above integral comes from $z\approx 0$, and we obtain
$$
m^2\simeq G(0)=\frac{1}{N_s}\sum_{{\bf k}} \gamma^2_{{\bf k}}
\delta(\xi_{\bf k}).
$$
Analogously, we have
$$
c=\frac{5 \Delta_0^2}{16} \int d z \; 
\frac{z^2}{(z^2+\Delta_0^2)^{\frac{7}{2}}} W(z)\simeq
\frac{W(0)}{12\,\Delta_0^2},
$$
where $W(z)=(1/N_s)\sum_{\bf k} ({\bf v}_{\bf k}^2/|\gamma_{\bf k}|)
\delta\left(z -\xi_{\bf k}/\gamma_{\bf k}\right)$ and $W(0)=V(0)$. As a 
consequence, according to the definition (\ref{average}), we can write
\begin{equation}
\xi_0^2= \frac{V(0)}{12\,\Delta_0^2\,G(0)}=\frac{v_F^2}{12\,\Delta_F^2},
\label{csi0g}
\end{equation}
which represent the generalization of the standard result for a continuum 
model, to the lattice case with arbitrary band dispersion 
$\xi_{\bf k}$, and for arbitrary symmetry $\gamma_{\bf k}$ of the gap 
parameter. 

In the $s$-wave case Eq. (\ref{csi0g}) reduces to 
$\xi_0^2=  v_F^2/12\Delta_0^2$, so that $\xi_{pair}$, Eq. (\ref{csis}), and 
$\xi_0$ differ only by the numerical factor, $\xi_{pair}/\xi_0=\sqrt{3/2}$. 
Both lengths are, in turn, proportional to the Pippard length scale 
$\xi_{e.m.}=  v_F/\pi\Delta_0$, which appears in the electromagnetic response 
function \cite{fetter,bcs}.

\section{The effect of a next-to-nearest-neighbor hopping}

Although the strong-coupling results obtained in Sec. III are specific
of the negative-$U$ Hubbard model with nearest neighbor hopping, we argue 
that they should not be dramatically modified by extensions of the original 
model. In particular, we discuss here in some detail the effect of a 
next-to-nearest-neighbor hopping term $t'$. It is known that such a term 
breaks the extended symmetry of the half-filled model with $t'=0$, making 
charge ordering unfavorable. At weak coupling this is the case, due to the 
spoiling of the perfect nesting of the Fermi surface as soon as $t'\neq 0$. 
Therefore, at weak coupling the superconducting state is favored at all 
doping down to $\delta=0$, and never becomes degenerate to the charge-ordered 
state. Since at small $U$ the coupling of the particle-particle and 
particle-hole channel is negligible, the weak-coupling results discussed in 
Sec. II are unaffected by the modification of the charge sector due to the  
$t'$ term. 

At strong coupling the extended symmetry is spoiled as well by the inclusion 
of $t'$ and again superconductivity is favored with respect to charge 
ordering, even at $\rho=1$. This can be more easily understood exploiting the 
spin analogue of the extended symmetry, which is achieved  by mapping the 
negative-$U$ Hubbard model onto a positive-$U$ model at half filling. We 
point out that, contrary to the case $t'=0$, in which the repulsive Hubbard 
model is recovered \cite{capone}, the transformed  $t'$ term acquires an extra 
sign which depends on the spin of the hopping fermion, 
$t'(-U) \rightarrow  \sigma t'(U)$. Subsequently, in the limit of large $U$, 
we map this positive-$U$ model onto an effective spin model which is the 
usual antiferromagnetic Heisenberg model with coupling $J\sim t^2/|U|$ for 
nearest-neighbor spins, but yields the non $SU(2)$-invariant coupling 
$$
J'\sum_{<<i,j>>} [S_i^z S_j^z - (S_i^x S_j^x + S_i^y S_j^y)], 
$$
with $J'\sim t'^2/|U|$, between spins located on next-to-nearest-neighbor 
sites $<<i,j>>$.
Evidently the magnetic order along the $z$-axis (which corresponds to charge 
ordering in the negative-$U$ Hubbard model) is frustrated due to the 
competition of the two antiferromagnetic couplings, whereas the magnetic order
on the $xy$-plane (which corresponds to superconductivity) is not frustrated 
due to the absence of interference between an antiferromagnetic 
nearest-neighbor and a ferromagnetic next-to-nearest-neighbor coupling. 
Indeed, the charge sector acquires a mass of order $t'^2/U$ at half filling. 
Nevertheless the results discussed in Sec. III are not dramatically 
modified, as long as the mass in the charge sector stays small near half 
filling, so that the strong peak at ${\bf q}={\bf Q}$ survives in 
$X_\rho({\bf q})$.

\end{multicols}


\begin{thebibliography}{99}
\bibitem{notadelta} We introduce the notation 
$\Delta_{\bf k}\equiv\Delta_0\gamma_{\bf k}$ for the ${\bf k}$-dependent
BCS gap. This should not be confused with $|\Delta|_q$ which is the
$q\equiv({\bf q},\Omega_m)$ component of the fluctuating amplitude $|\Delta|$
of the order parameter.
\bibitem{notamu} As it is customary, our chemical potential $\mu$ includes
the Hartree shift $\mu=\tilde\mu+(U/2)\rho$, where $\tilde \mu$ is the true
chemical potential.
\end{thebibliography}
\end{document}